\documentclass[12pt]{article}
\usepackage{amsmath,amssymb}
\usepackage{bm}
\usepackage{color}
\usepackage{cite}
\usepackage{mathtools}

\usepackage{multirow}

\begin{document}

\title{
\begin{flushright}
\ \\*[-80pt]
\begin{minipage}{0.25\linewidth}
\normalsize
EPHOU-22-007\\
KEK-TH-2407\\
KYUSHU-HET-238 \\*[50pt]
\end{minipage}
\end{flushright}
{\Large \bf
Modular symmetry \\of soft supersymmetry breaking terms
\\*[20pt]}}

\author{
Shota Kikuchi$^{1}$,
Tatsuo Kobayashi$^{1}$,
Kaito Nasu$^{1}$, \\
 Hajime Otsuka$^{2}$, 
Shohei Takada$^{1}$, 
and
~Hikaru Uchida$^{1}$
\\*[20pt]
\centerline{
\begin{minipage}{\linewidth}
\begin{center}
$^{1}${\it \normalsize
Department of Physics, Hokkaido University, Sapporo 060-0810, Japan} \\*[5pt]
$^{2}${\it \normalsize
KEK Theory Center, Institute of Particle and Nuclear Studies, 1-1 Oho, Tsukuba, Ibaraki 305-0801, Japan} \\*[5pt]
$^{3}${\it \normalsize
Department of Physics, Kyushu University, 744 Motooka, Nishi-ku, Fukuoka 819-0395, Japan} \\*[5pt]
\end{center}
\end{minipage}}
\\*[50pt]}

\date{
\centerline{\small \bf Abstract}
\begin{minipage}{0.9\linewidth}
\medskip
\medskip
\small
We study the modular symmetry of soft supersymmetry breaking terms.
Soft scalar masses and $A$-term coefficients are invariant under the modular symmetry 
when we regard $F$-term as a spurion with the modular weight $-2$.
Their flavor structure is determined by the same symmetry as Yukawa couplings, i.e., fermion masses. 
The modular symmetric behavior of $\mu$-term and $B$-term depends on how the $\mu$-term is generated. 
\end{minipage}
}

\begin{titlepage}
\maketitle
\thispagestyle{empty}
\end{titlepage}

\newpage


\section{Introduction}
\label{Intro}

Superstring theory is a promising candidate for a unified theory of all the interactions including gravity, 
matter such as quarks and leptons, and the Higgs field.
One needs compactification of six extra dimensions to realize four-dimensional (4D) theory because 
superstring theory is consistent in ten dimensions.
It is important to study properties of 4D low energy effective field theory below the compactification scale 
in order to derive particle physics from superstring theory.
The compactification scale is expected to be quite high compared with the weak scale.
Symmetries would be quite useful to bridge superstring theory and particle physics at the low energy scale.
For example, the flavor symmetries originated from geometrical symmetries and stringy coupling selection rules 
were studied in Refs. \cite{Kobayashi:2004ya,Kobayashi:2006wq,Abe:2009vi}.

Recently, the modular symmetry has been studied extensively from both top-down and bottom-up approaches.
The modular symmetry is a geometrical symmetry of compact spaces such as the torus $T^2$ and 
the orbifold $T^2/Z_2$.
The 4D supergravity  theory derived from string theory also has 
the modular symmetry \cite{Ferrara:1989bc}.
Furthermore, zero-modes transform each other under the modular symmetry in string-derived low energy effective field theories \cite{Ferrara:1989qb,Lerche:1989cs,Lauer:1990tm,Kobayashi:2018rad,Kobayashi:2018bff,Ohki:2020bpo,Kikuchi:2020frp,Kikuchi:2020nxn,
Kikuchi:2021ogn,Almumin:2021fbk,Baur:2019iai,Nilles:2020kgo,Baur:2020jwc,Nilles:2020gvu}.
That is, the modular symmetry is a flavor symmetry.
Finite modular groups include $S_3$, $A_4$, $S_4$, and $A_5$ groups \cite{deAdelhartToorop:2011re}.

On the other hand, in the bottom-up approach, 
the above non-Abelian discrete flavor symmetries like $S_N$ and $A_N$ were often used in model building 
to derive the flavor structure of the quarks and leptons such as fermion masses, mixing angles, and CP phases 
\cite{
	Altarelli:2010gt,Ishimori:2010au,Kobayashi:2022moq,Hernandez:2012ra,
	King:2013eh,King:2014nza}.
That inspired the model building with the modular flavor symmetries, 
and many models have been studied in the bottom-up approach.
(See e.g., Refs.~\cite{Feruglio:2017spp,Kobayashi:2018vbk,Penedo:2018nmg,Novichkov:2018nkm,Criado:2018thu,
Kobayashi:2018scp,
Ding:2019zxk,Novichkov:2018ovf,
Kobayashi:2019mna,Wang:2019ovr,Ding:2019xna,
Liu:2019khw,Chen:2020udk,Novichkov:2020eep,Liu:2020akv,
deMedeirosVarzielas:2019cyj,
  	Asaka:2019vev,Ding:2020msi,Asaka:2020tmo,deAnda:2018ecu,Kobayashi:2019rzp,Novichkov:2018yse,Kobayashi:2018wkl,Okada:2018yrn,Okada:2019uoy,Nomura:2019jxj, Okada:2019xqk,
  	Nomura:2019yft,Nomura:2019lnr,Criado:2019tzk,
  	King:2019vhv,Gui-JunDing:2019wap,deMedeirosVarzielas:2020kji,Zhang:2019ngf,Nomura:2019xsb,Kobayashi:2019gtp,Lu:2019vgm,Wang:2019xbo,King:2020qaj,Abbas:2020qzc,Okada:2020oxh,Okada:2020dmb,Ding:2020yen,Okada:2020rjb,Okada:2020ukr,Nagao:2020azf,Wang:2020lxk,
  	Okada:2020brs,Yao:2020qyy}.)\footnote{See for more recent references Ref.~\cite{Kikuchi:2022txy}.}

In addition, a generic six-dimensional compact space has more than one modulus, and they would have 
larger geometrical symmetries.
For example, Calabi-Yau compactifications have many moduli.
The 4D low energy effective field theory has a larger symplectic modular symmetry \cite{Strominger:1990pd,Candelas:1990pi,Ishiguro:2020nuf,Ishiguro:2021ccl}.

One of the important issues in this context is supersymmetry (SUSY) breaking to discuss low energy particle physics 
from string-derived supergravity theory.
Studies on SUSY breaking effects are important in the modular flavor symmetric models of 
the bottom-up approach \cite{Kobayashi:2021jqu,Tanimoto:2021ehw,Otsuka:2022rak} 
because most of those models are constructed within the SUSY framework.
The pattern of soft SUSY breaking terms such as sfermion masses and $A$-terms depends on the source of SUSY breaking.
If the SUSY breaking source is (trivial) singlet under the modular symmetry, 
soft SUSY breaking terms would be obviously invariant under the modular symmetry.
Thus, our purpose is to study soft SUSY breaking terms caused by the $F$-term of the modulus, 
which transforms non-trivially under the modular symmetry.
That is, we assume the so-called moduli mediated SUSY breaking 
\cite{Kaplunovsky:1993rd,Brignole:1993dj,Kobayashi:1994eh,Brignole:1995fb,Ibanez:1998rf}
without specifying SUSY breaking potential of moduli.
Then, we study the behavior of soft SUSY breaking terms under the modular symmetry.

This paper is organized as follows.
In section 2, we briefly review the modular symmetry.
In section 3, we study the behavior of soft SUSY breaking terms under the modular symmetry. Section 4 is devoted to the conclusion.

\section{Modular symmetry}
\label{sec:mod-symm}

Here, we give a brief review on the modular symmetry.
Under the modular symmetry, the modulus $\tau$ transforms 
\begin{align}
\tau \to \gamma \tau = \frac{a\tau + b}{c \tau + d},
\end{align}
where 
\begin{align}
\gamma=
\begin{pmatrix}
a & b \\
c & d
\end{pmatrix},
\end{align}
is an element of  the $SL(2,\mathbb{Z})$ group satisfying 
$ad-bc=1$ for $a,b,c,d=$ integers.
The modular group is generated by $S$ and $T$:
\begin{align}
S=
\begin{pmatrix}
0 & 1 \\ -1 &0
\end{pmatrix}, 
\qquad 
T=
\begin{pmatrix}
1 & 1 \\ 0 & 1
\end{pmatrix}.
\end{align}
They transform the modulus $\tau$ as 
\begin{align}
S~:~\tau \to -1/\tau, \notag \\
T~:~\tau \to \tau+1.
\end{align}

The modular forms  are holomorphic functions $f_i(\tau)$, which 
transform
\begin{align}
f_i(\tau) \to f_i(\gamma \tau)=\rho(\gamma)_{ij}(c\tau + d)^{k_i}f_j(\tau),
\end{align}
under the modular symmetry. Here, $k_i$ and the factor $(c\tau + d)^{k_i}$ are respectively called as the modular weight and the automorphy factor, and $\rho(\gamma)_{ij}$ is a unitary matrix.
They also transform 
\begin{align}
f_i(\tau) \to f_i(\gamma \tau)=(c\tau + d)^{k_i}f_i(\tau),
\end{align}
for $\gamma$ in a subgroup of the modular group such as 
congruence subgroups.

In addition, we also introduce derivatives of modular forms, ${df_i}/{d\tau}$.
They transform 
\begin{align}
\label{eq:form-derivative}
\frac{df_i(\tau)}{d \tau} \to \rho(\gamma)_{ij}\left((c\tau + d)^{k_i+2}\frac{df_j(\tau)}{d \tau}  + ck_i(c\tau+d)^{k_i+1} f_j(\tau) 
\right).
\end{align}
Since the same unitary matrix $\rho(\gamma)_{ij}$ appears in the above form, we define 
\begin{align}
D_\tau f_i(\tau) \equiv \frac{df_i(\tau)}{d \tau} -i\frac{k_i}{2{\rm Im}(\tau)}f_i(\tau).
\end{align}
It means that the covariant derivative of modular forms transform as the modular form with the weight $k_i+2$,
\begin{align}
\label{eq:covariant-derivative}
D_\tau f_i(\tau)~\to~\rho(\gamma)_{ij}(c\tau +d)^{k_i+2}D_\tau f_j(\tau).
\end{align}

In general, we can define
\begin{align}
\tilde D_\tau f_i(\tau) \equiv \frac{df_i(\tau)}{d \tau} -f_i(\tau) \frac{d }{d \tau} \ln (g(\tau, \bar \tau)),
\end{align}
where $g(\tau, \bar \tau)$ has the holomorphic modular weight $-k_i$ and arbitrary anti-holomorphic modular weight.
Then, the modular symmetric behavior is the same as Eq. (\ref{eq:covariant-derivative}).
At any rate, the above definition of covariant derivative $D_\tau f_i(\tau)$ is convenient in what follows.

\section{Soft SUSY breaking terms}
\label{sec:SUSYbreaking}

In this section, we study soft SUSY breaking terms induced by the moduli $F$-terms 
within the framework of supergravity theory.

\subsection{SUSY breaking terms in the moduli mediation}

We denote the superfield including the modulus $\tau$ as 
\begin{align}
\label{eq:superfield}
U = \tau +\theta^2 F^U.
\end{align}
Here, we have omitted the fermionic superpartner of $\tau$ 
because it is irrelevant to the following discussions.
In general, there are superfields $M$ including other moduli fields.

Let us assume that the superfield $U$ transforms 
\begin{align}
U ~\to~ \frac{aU + b}{c U + d},
\end{align}
under the modular symmetry in the same way as $\tau$.
The $F$-term $F^U$ transforms
\begin{align}
\label{eq:FU-modular}
F^U ~\to~ \frac{F^U}{(c\tau + d)^{2}}.
\end{align}
Thus, the $F$-term $F^U$ has the modular weight $-2$. 
Note that the other moduli $M$ are invariant under the modular symmetry of $\tau$.
The $F$-term $F^U$ appears in Lagrangian as
\begin{align}
K_{U\bar U}|F^U|^2=\frac{|F^U|^2}{(i(\bar U-U))^2},
\end{align}
where $K_{U \bar U}$ is the K\"ahler metric of $U$.
This is consistent with the above modular transformation (\ref{eq:FU-modular}).

We study a supergravity theory with the following K\"ahler potential\footnote{Throughout this paper, we adopt the reduced Planck mass unit $M_{\rm Pl}=1$.}:
\begin{eqnarray}\label{kahler}
K & =& K_0(U,\bar U, M,\bar M)+
K^{\rm matter} \,, \nonumber \\
K_0(\tau,M) &=& -\ln(i(\bar U -U)) + K(M,\bar M)\,,  \\
K^{\rm matter} &=& K_{i\bar i}|\Phi^i|^2, \qquad K_{i\bar i}= \frac{1}{(i(\bar U -U))^{k_i}} \nonumber \,,
\end{eqnarray}
and the superpotential:
\begin{eqnarray}
W= W_0(U,M)+Y_{ijk}(U)\Phi_i \Phi_j \Phi_k  +\cdots \,.
\label{super}
\end{eqnarray}
Here, $\Phi_i$ denote chiral matter superfields and the ellipsis corresponds 
to the higher-order terms of $\Phi_i$.\footnote{Moduli superpotential is assumed to be modular invariant, but it would be realized in e.g., 
flux compactifications of the string theory \cite{Kobayashi:2020hoc}.} 
Under the modular symmetry, the matter superfields transform 
\begin{align}
\Phi^i~\to~\frac{\rho(\gamma)_{ij}}{(cU+d)^{k_i}}\Phi^j.
\end{align}
They have the modular weight $-k_i$.
The matrix $\rho(\gamma)_{ij}$ represents the modular flavor symmetry.
The K\"ahler metric $K_{i\bar i}$ is diagonal in many string-derived models as well as 
models from the bottom-up approach because 
they have different quantum numbers each other for $\rho(\gamma)_{ij}$ and other symmetries. (See for the off-diagonal case in section \ref{sec:multi}.)  
We assume that the Yukawa coupling $Y_{ijk}$ depends only on the modulus $U$.
In addition, we consider the gauge kinetic functions independent of $U$, as realized in the effective action of Type IIB string theory. \footnote{Even at the loop level, loop corrections are characterized by the modular invariant function \cite{Blumenhagen:2006ci}.}
The gaugino masses are proportional to $F^M$, which are invariant under the modular symmetry.

The combination $e^K|W|^2$ of $K$ and $W$ must be invariant under the modular symmetry.
That is, the superpotential $W$ has the modular weight $-1$.
Indeed, the gravitino mass $m_{3/2}$ is obtained as 
\begin{align}
m^2_{3/2}=\langle e^{K_0} |W_0|^2 \rangle,
\end{align}
which is invariant under the modular symmetry. 
Furthermore, the $F$-term scalar potential $V$ is written by 
\begin{align}
V=e^K \left(K^{I \bar J}(W_I+K_IW)(\bar W_{\bar J} + K_{\bar J} \bar W)-3
|W|^2\right),
\end{align}
where the index $I$ includes the moduli $X$ and matter fields $\Phi^i$, 
and $K^{I \bar J}$ is the inverse of the K\"ahler metric $K_{I \bar J}$.
The scalar potential is invariant under the modular symmetry.
For example, we find that $(W_I+K_IW)$ transforms
\begin{align}
(W_\tau+K_\tau W)~\to~(c\tau+d)(W_\tau+K_\tau W),
\end{align}
because of Eq.(\ref{eq:covariant-derivative}), and 
\begin{align}
(W_M+K_M W)~\to~(c\tau+d)^{-1}(W_M+K_M W),
\end{align}
for other fields.
We expect that soft SUSY breaking terms would be invariant under the modular symmetry 
when we regard $F^U$ as the spurion, which transforms as in Eq. (\ref{eq:FU-modular}).
We examine their behavior by explicit computations of soft SUSY breaking terms caused by $F^U$.

In addition, we have the $D$-term scalar potential.
For example, for $U(1)$ gauge theory, the $D$-term potential can be written by 
\begin{align}
V_D=\frac12 g^2 D^2, \qquad D=\sum q_i K_{i\bar i} |\phi^i|^2,
\end{align}
where $g$ is the gauge coupling, $\phi^i$ is the scalar component of chiral superfield $\Phi^i$, and $q_i$ denotes the $U(1)$ charge of $\Phi^i$.
We find that $D$ is invariant under the modular symmetry, and 
scalar masses induced by the $D$-term term, $m_{i\bar i(D)}^2=g^2Dq_i$ is invariant under the modular symmetry.

Suppose that the $F$-terms of moduli develop their vacuum expectation values and 
break SUSY.
Then, we calculate soft SUSY breaking terms.

First, we study sfermon masses, which are written by \cite{Kaplunovsky:1993rd}
\begin{eqnarray}
\tilde m_i^2= m_{3/2}^2-\sum_{X=U, M} |F^X|^2 \partial_X \partial_{\bar X}\ln K_{i \bar i}\,,
\end{eqnarray}
in the canonically normalized basis.
By using the above K\"ahler metric $K_{i\bar i}$ of matter fields, we obtain 
\begin{eqnarray}
\tilde m_i^2= m_{3/2}^2-k_i \frac{|F^U|^2}{(2{\rm Im}(\tau))^2} \,.
\end{eqnarray}
It is found that the sfermion masses are invariant under the modular symmetry 
because the gravitino mass $m_{3/2}$ is invariant and $F^U$ has the modular weight $-2$.

Next, we study the $A$-terms, which appear in SUSY breaking Lagrangian,
\begin{align}
{\cal L}_{\rm A} = h_{ijk}\phi^i\phi^j\phi^k + {\rm h.c.}, 
\end{align}
in the non-canonically normalized basis of $\phi^i$.
We use the convention $h_{ijk}=Y_{ijk}A_{ijk}$.
Then, the $A$-term coefficient $A_{ijk}$ is written by \cite{Kaplunovsky:1993rd}
\begin{eqnarray}
A_{ijk} =A_i+A_j+A_k -\sum_{X=U, M} \frac{F^X}{Y_{ijk}} \partial_X Y_{ijk}\,, 
\end{eqnarray}
with 
\begin{eqnarray}
A_i = \sum_X F^X \partial_X \ln e^{-K_0/3}K_{i\bar i}\,.
\end{eqnarray}

The Yukawa couplings in supergravity theory may be  related with Yukawa couplings 
$\hat Y_{ijk}$ in global SUSY theory 
as  $|\hat Y_{ijk}|^2=e^{K_0}|Y_{ijk}|^2$ \cite{Kaplunovsky:1993rd}, 
but we include the factor $e^{K_0}$ in the kinetic term and redefine the following 
kinetic factor,
\begin{align}
\hat K_{i \bar \bar i}=e^{-K_0/3}K_{i \bar i},
\end{align}
in order to preserve the holomorphy of the Yukawa couplings $Y_{ijk}$.

By using the explicit form of K\"ahler metric and Yukawa coupling, we obtain 
\begin{eqnarray}
A_{ijk}&=&A^0_{(M)}+A_{ijk}^0+A'_{ijk}, \nonumber \\
A_{ijk}^0&=& i(k_i+k_j+k_k -1)\frac{ F^U}{2{\rm Im}(\tau)}, \qquad 
A'_{ijk}=-\frac{F^U}{Y_{ijk}}\frac{dY_{ijk}(\tau)}{d \tau} \, ,
\label{Aterm}
\end{eqnarray}
where $A^0_{(M)}$ is the universal contribution due to other moduli $M$.
It is obvious that $A^0_{(M)}$ is invariant under the modular symmetry; $Y_{ijk}A^0_{(M)}$ transforms in the same way as $Y_{ijk}$.
We examine the modular transformation of the other terms,
\begin{align}
h^0_{ijk}=Y_{ijk}A^0_{ijk}, \qquad h'_{ijk}=Y_{ijk}A'_{ijk}=-F^U\frac{dY_{ijk}(\tau)}{d \tau}.
\end{align}
We assume that the Yukawa coupling $Y_{ijk}$ transforms as
\begin{align}
Y_{ijk}(\gamma \tau)=\rho(\gamma)_{(ijk)(\ell mn)}(c\tau + d)^{k_Y}Y_{\ell mn}(\tau),
\end{align}
where $k_Y$ is the modular weight of Yukawa coupling.
Note that it must satisfy the following relation: 
\begin{align}
\label{eq:weight-relation}
k_Y=k_i+k_j+k_k-1,
\end{align}
because $e^K|W|^2$ and $e^{K_0}|Y_{ijk}\Phi^i\Phi^j \Phi^k|^2$ are invariant under the modular symmetry.
By using Eq.~(\ref{eq:form-derivative}), we find the transformation of $h'_{ijk}$ as 
\begin{align}
h'_{ijk}~\to~&\rho(\gamma)_{(ijk)(\ell m n)} \frac{- F^U}{(c\tau + d)^2}\left((c\tau + d)^{k_Y+2}\frac{dY_{\ell m n}(\tau)}{d \tau}  + ck_Y(c\tau+d)^{k_Y+1} Y_{\ell m n}(\tau) \right) \notag \\
&=\rho(\gamma)_{(ijk)(\ell m n)} (c\tau + d)^{k_Y}h'_{\ell mn}+\rho(\gamma)_{(ijk)(\ell mn)} \Delta h'_{\ell mn},
\end{align}
where 
\begin{align}
\Delta h'_{\ell mn}=-ck_Y(c\tau + d)^{k_Y-1}Y_{\ell mn}(\tau)F^U.
\end{align}
If $\Delta h'_{\ell mn}=0$, the transformation behavior of $h'_{ijk}$ by itself is the same as $Y_{ijk}$.
It is not the case except $k_Y=0$.
However, $\Delta h'_{\ell m n}$ is canceled by the transformation of $h^0_{ijk}$.
Indeed, $A^0_{ijk}$ transforms
\begin{align}
A^0_{ijk}~\to~k_Y\frac{iF^U}{2{\rm Im}(\tau)}\frac{c \bar \tau +d}{c \tau +d} 
=A^0_{ijk}+\frac{ck_Y}{c\tau +d}F^U,
\end{align}
where we have used Eq.~(\ref{eq:weight-relation}).
Thus, we find the following transformation under the modular symmetry:
\begin{align}
h^0_{ijk}~\to~\rho(\gamma)_{(ijk)(\ell m n)}(c\tau +d)^{k_Y}h^0_{\ell m n}+\rho(\gamma)_{(ijk)(\ell m n)}ck_Y(c\tau +d)^{k_Y-1} Y_{\ell m n}(\tau) F^U.
\end{align}
The second term cancels with $\rho(\gamma)_{(ijk)(\ell m n)}\Delta h'_{\ell m n}$.
As a result, $Y_{ijk}(A^0_{ijk}+A'_{ijk})$ transforms in the same way as $Y_{ijk}$.
That is because $(A^0_{ijk}+A'_{ijk})$ behaves like a covariant derivative for the modular symmetry.
Indeed, we can write
\begin{align}
Y_{ijk}(A^0_{ijk}+A'_{ijk})&=-F^U\left[\frac{dY_{ijk}}{d\tau} + \frac{d}{d \tau}\ln \left[ e^{K_0}/(K_{i\bar i} K_{j \bar j} K_{k \bar k})\right]
\right], \notag \\
&= -F^U\left[\frac{dY_{ijk}}{d\tau} -i \frac{k_Y}{2 {\rm Im}(\tau)}Y_{ijk} \right] \notag \\
&=-F^UD_\tau Y_{ijk},
\end{align}
although we calculated $Y_{ijk}A^0_{ijk}$ and  $Y_{ijk}A'_{ijk}$  explicitly and separately in the above.

\subsection{Higgs sector}

Let us study the Higgs sector, the $\mu$-term and $B$-term.
Superstring theory does not lead to mass terms of light modes at the tree level.
One possibility to generate the $\mu$-term is due to non-perturbative effects.
Another way is the Giudice-Masiero mechanism \cite{Giudice:1988yz}.
First, we study the latter mechanism.

In addition to the above K\"ahler potential (\ref{kahler}), we add the following term:
\begin{align}
\Delta K_{GM}=H(X,\bar X)H_uH_d + {\rm h.c.}, 
\end{align}
where $H_u$ and $H_d$ are Higgs superfields of the up- and down-sectors, respectively.
Indeed,  a certain type of heterotic orbifold models leads to 
the following form of matter K\"ahler potential \cite{Brignole:1995fb}, 
\begin{align}
K^{\rm matter}=\frac{|H_u+\bar H_d|^2}{-(\bar X_1 -X_1)(\bar X_2 -X_2)},
\end{align}
which corresponds to 
\begin{align}
K_{H_{u,d}\bar H_{u,d}}=H(X,\bar X)= \frac{1}{-(\bar X_1 -X_1)(\bar X_2 -X_2)}.
\end{align}
Then, in the non-canonically normalized basis, the $\mu$-term is written by \cite{Kaplunovsky:1993rd}
\begin{align}
\mu= m_{3/2}H(X,\bar X)-\sum_X F^X\partial_XH(X,\bar X).
\end{align}
When $X_{1,2}=M$ other than $U$, the $\mu$-term is invariant under the modular symmetry.
The so-called $B$-term in Lagrangian is written  as
\begin{align}
{\cal L}_B=bh_uh_d + {\rm h.c.},
\end{align}
where $h_{u,d}$ are the scalar components of $H_{u,d}$, respectively.
When the $\mu$-term is generated by the Giudice-Masiero 
mechanism not including $U$, i.e., $X_{1,2}\neq U$, the $B$-term is independent of $U$, and it is also invariant 
under the modular symmetry of $U$. 
(See for a generic formula of the $B$-term in Ref.~\cite{Kaplunovsky:1993rd}.)

On the other hand, when $X_1=U$, the $\mu$-term does not seem to be invariant under the modular symmetry.
This is because the additional term in the K\"ahler potential,
 \begin{align}
\label{eq:K-GM}
\Delta K_{GM}=\frac{H_uH_d}{-(\bar U-U)(\bar X_2 -X_2)} + {\rm h.c.},
\end{align}
is not consistent with the modular symmetry of $U$.
Although we may assign the modular transformation such that 
$H_d$ transforms in an anti-holomorphic way, such an assignment is not 
consistent with the holomorphic Yukawa coupling in the superpotential\footnote{If we can define the transformation, $H_{u,d} \to H_{u,d}/|c\tau+d|$, 
$\Delta K_{GM}$ in Eq.~(\ref{eq:K-GM}) may be consistent.
Such a real scale transformation may remain as a symmetry.}.

Alternatively, the $\mu$-term may be induced by non-perturbative effects.
Here, we assume such a scenario and the following superpotential term:
\begin{align}
W_{\rm np}=\mu (\tau) H_u H_d,
\end{align}
is induced.
If such a term is also modular symmetric, the weight of $\mu(\tau)$, $k_\mu$ must satisfy 
\begin{align}
\label{eq:weight-mu}
k_{\mu}=k_{H_u}+k_{H_d}-1,
\end{align}
because $e^{K_0}|\mu(\tau) H_u H_d|^2$ is modular invariant.
In addition, $\mu(\tau)$ is singlet under the modular symmetry in the minimal supersymmetric standard model.
Thus, $\mu(\tau)$ transforms under the modular symmetry,
\begin{align}
\mu(\tau)~\to~(c\tau + d)^{k_\mu} \mu(\tau).
\end{align}
Note that the global SUSY $\mu$-term is defined as 
$|\hat \mu|^2= e^{K_0}|\mu|^2$ \cite{Kaplunovsky:1993rd}.
Such a $\mu$-term in the global SUSY with canonically normalized basis 
is invariant under the modular symmetry.
However, non-perturbative effects may break the modular symmetry, and 
$e^{K_0}|\mu(\tau) H_u H_d|^2$ is not invariant.
In such a case, the $\mu$-term is not invariant \footnote{
In Ref.~\cite{Kikuchi:2022bkn}, right-handed neutrino mass terms induced D-brane instanton effects were studied.
In those models, the modular weight of neutrino mass terms does not match with 
other terms in Lagrangian at the tree level.}.
Thus, it depends on whether non-perturbative effects break the modular symmetry.
If non-perturbative effects preserve the modular symmetry and the $\mu$-term is invariant under the modular symmetry, 
the $B$-term coefficient, $B=b/\mu$ may be invariant under the modular symmetry in a way similar to 
the $A$-term coefficient, $A_{ijk}$.

Now, we write down the Higgs potential $V_H$,
\begin{align}
V_H=&(|\mu|^2+ m^2_{Hu})(|\tilde h_u^+|^2+|\tilde h_u^0|^2)+ (|\mu|^2+m^2_{Hd})(|\tilde h_d^0|^2+|\tilde h_d^-|^2) \notag \\
& +\mu B(\tilde h_u^+\tilde h_d^- -\tilde h_u^0\tilde h_d^0) + {\rm h.c.}\\
&+\frac{g^2}{2}|\tilde h_u^+\tilde h_d^{0*}+\tilde h_u^0 \tilde h_d^{-*}|^2 
+\frac18(g^2+g'^2) (|\tilde h_u^+|^2+|\tilde h_u^0|^2-|\tilde h_d^0|^2-|\tilde h_u^-|^2)^2,  \notag
\end{align}
for canonically normalized fields, 
\begin{align}
\tilde h_u=
\begin{pmatrix}
\tilde h_u^+ \\ \tilde h_u^0
\end{pmatrix},
\qquad  \tilde h_d=
\begin{pmatrix}
\tilde h_d^0 \\ \tilde h_d^-
\end{pmatrix}.
\end{align}
The $SU(2)$ and $U(1)_Y$ gauge couplings, $g$ and $g'$, are invariant under the modular symmetry.
Also, all of the quadratic terms are invariant if the $\mu$-term is generated in a way consistent with 
the modular symmetry, e.g., the Giudice-Masiero mechanism with $H(M,\bar M)$ for $M\neq U$.
Thus, the whole of the Higgs sector can be invariant.


\subsection{Sfermion sector}
\label{sec:sfermion}

The  mass terms of left- and right-handed up-sector squarks,  $\phi_{Li}$ and $\phi_{Ri}$ 
in the canonically normalized basis can be written by 
\begin{align}
{\cal L}_{\rm masses}=-(\phi^*_L,\phi^*_R)~M^2_{ij}
\begin{pmatrix}
\phi_L \\ \phi_R
\end{pmatrix},
\end{align}
with
\begin{align}
\label{eq:squark-mass}
M^2_{ij}=
\begin{pmatrix}
m^2_{ij(LL)} +y^\dag_{ikH_u} y_{kjH_u} v_u^2 & (Ay)^\dag_{ijH_u}v_u-\mu y^\dag_{ijH_u}v_d \\
(Ay)_{ijH_u}v_u-\mu y_{ijH_u}v_d & m^2_{ij(RR)} +y^\dag_{ikH_u}y_{kjH_u} v_u^2
\end{pmatrix},
\end{align}
where $m^2_{ij(LL)}$ and $m^2_{ij(RR)}$ are left- and right-handed scalar masses, 
$y_{ijH_u}$ is the Yukawa couplings in the canonically normalized basis, and 
$v_{u,d}$ denote vacuum expectation values of $\tilde h_{u,d}$.
The soft scalar masses, $m^2_{ij(LL)}$ and $m^2_{ij(RR)}$ and $A$-term coefficients $A_{ijH_u}$ are 
invariant under the modular symmetry as studied in the previous section.
The modular transformation behavior of the  $\mu$-term depends on its generation mechanism.
If its generation mechanism is consistent with the modular symmetry, e.g., the 
Giudice-Masiero mechanism, the $\mu$-term is also invariant under the modular symmetry.

The soft scalar mass matrices  $m^2_{ij(LL)}$ and $m^2_{ij(RR)}$  are diagonal as a consequence of diagonal K\"ahler metric of matter fields.
Moreover,  if three generations of quarks have the same modular weight $k_i$, 
the soft scalar mass matrices  $m^2_{ij(LL)}$ and $m^2_{ij(RR)}$  are proportional to 
the $(3\times 3)$ unit matrix.
Such a flavor structure can be realized when three generations are irreducible representations of 
$\rho(\gamma)_{ij}$.
In this case, the flavor structure in the $M^2_{ij}$ in Eq. (\ref{eq:squark-mass}) are determined by 
$\rho(\gamma)$, which is the same as the flavor structure of the Yukawa matrix in the quark sector.
Similarly, we can study mass terms of the down-sector squarks and sleptons.

Eigenvalues of $M^2_{ij}$ with the Yukawa couplings $y_{ijH_u}$ are the same as 
those with $(\rho(\gamma)y)_{ijH_u}$ because that is just a change of the representation basis of the modular flavor symmetry. 
Thus, these eigenvalues are invariant under the modular flavor symmetry. 
The same is true for the down-sector squarks and sleptons.
We denote eigenvalues of these sfermion masses squared 
$m^2_{\tilde f}(v_u,v_d)$ as functions of vacuum expectation values 
as well as masses squared of their fermionic superpartners 
$m^2_f(v_u,v_d)$.
The neutral Higgs one-loop effective potential can be written by \cite{Carena:1995wu}
\begin{align}
V^{(1)}(\tilde h_u^0=v_u, \tilde h^0_d=v_d)=\frac{3}{32\pi^2}\left[ \sum_{\tilde f} m^4_{\tilde f} \left(
\ln \frac{m^2_{\tilde f}}{Q^2}       -\frac32 \right)
-2 \sum_f m^4_f\left( \ln \frac{m^2_f}{Q^2} -\frac32 \right) \right],
\end{align}
by replacing the vacuum expectation values with $\tilde h_u^0=v_u$ and $ \tilde h^0_d=v_d$.
That includes large threshold corrections on the Higgs quartic coupling 
due to the top Yukawa coupling \cite{Okada:1990vk,Haber:1990aw,Ellis:1990nz} and the left-right stop mixing \cite{Carena:1995wu}.
These corrections and the one-loop effective potential by itself are 
invariant under the modular flavor symmetry because eigenvalues of masses squared 
are invariant.

We also give a comment on the SUSY threshold corrections on Yukawa couplings.
For example, such threshold corrections on the down-sector Yukawa couplings are induced by 
the loop effects, including the up-sector $A$-terms \cite{Hall:1993gn,Hempfling:1993kv}.
Thus, such threshold corrections are controlled by the modular flavor symmetry, $\rho(\gamma)$.

\subsection{Multi moduli theory: symplectic modular symmetry}
\label{sec:multi}

Generic compact space has more than one modulus, $U^i$, and 
there is a larger modular symmetry.
Each of $U^i$ has $SL(2,\mathbb{Z})_i$ symmetry, 
and it can be discussed in a way similar to the previous discussion.
In addition, we have a permutation symmetry among $U^i$.
For concreteness, we briefly study 4D low energy effective theory of 
heterotic string theory on Calabi-Yau compactification with the 
standard embedding, where the complex structure moduli and ${\bf \overline{27}}$ matter fields of $E_6$ are connected on one-to-one correspondence. 
The following discussion is also applicable to the K\"ahler moduli sector, corresponding to the ${\bf 27}$ matter fields of $E_6$. 

In this context, one can discuss the symplectic modular flavor symmetry thanks to the identification of the moduli with matter fields \cite{Ishiguro:2021ccl}. Indeed, 
the Yukawa couplings of ${\bf 27}$ and ${\bf \overline{27}}$ are determined by the 
third derivatives of the prepotential in the corresponding moduli. After a brief 
review of the symplectic modular symmetry, we discuss the symplectic modular symmetry of soft SUSY breaking terms.

A Calabi-Yau threefold ${\cal M}$ has a single holomorphic three-form, $\Omega$. 
which can be expanded by the symplectic basis $(\alpha_I,\beta^I)$ of $H^3({\cal M},\mathbb{Z})$, 
\begin{align}
\Omega =\sum_{I=0}^{h^{2,1}}(X^I \alpha_I -{\cal F}_I \beta^I).
\end{align}
The basis satisfies 
\begin{align}
\int_{\cal M}\alpha_I \wedge \beta^J = \delta^J_I, \qquad \int_{\cal M}\beta^J \wedge \alpha_I = -\delta^J_I.
\end{align}
Here, ${\cal F}_I$ is obtained by the prepotential ${\cal F}$ as ${\cal F}_I=\partial_I {\cal F}$, 
and the prepotential is written in the large complex structure regime by 
\begin{align}
{\cal F}=\frac{\kappa_{ijk}}{3 !}\frac{X^i X^j X^k}{X^0},
\end{align}
where $\kappa_{ijk}$ denotes the triple intersection numbers.\footnote{Here, we omit the geometric corrections originating from instanton effects in the topological A-model.}

We consider the basis transformation
\begin{align}
\begin{pmatrix}
\alpha_I \\ \beta^I
\end{pmatrix}
~\to~
\begin{pmatrix}
a & b\\ c & d
\end{pmatrix}
\begin{pmatrix}
\alpha_I \\ \beta^I
\end{pmatrix},
\end{align}
where
\begin{align}
\begin{pmatrix}
a & b\\ c & d
\end{pmatrix}
\in Sp(2h^{2,1}+2,\mathbb{Z}).
\end{align}
Then, $(X^I, {\cal F}_I)^T$ must transform 
\begin{align}
\begin{pmatrix}
X^I \\ {\cal F}_I
\end{pmatrix}
~\to~
\begin{pmatrix}
\tilde X^I \\ \tilde {\cal F}_I
\end{pmatrix}
=
\begin{pmatrix}
d & c\\ b & a
\end{pmatrix}
\begin{pmatrix}
X^I \\ {\cal F}_I
\end{pmatrix},
\end{align}
for $\Omega$ to be invariant, and 
$X^I$ transform
\begin{align}
X^I~\to~\tilde X^I=(c{\cal F}+d)^I_JX^J=\frac{\partial \tilde X^I}{\partial X^J}X^J.
\end{align}

The complex structure moduli are defined by 
\begin{align}
u^i=\frac{X^i}{X^0},
\end{align}
and they transform 
\begin{align}
u^i~\to~\tilde u^i=(c{\cal F}+d)^i_j\frac{u^j}{\tilde X^0}=\frac{\partial \tilde X^i}{\partial X^j}\frac{u^j}{\tilde X^0}.
\end{align}
Similar to Eq.(\ref{eq:superfield}), we extend the moduli $u^i$ to the superfields $U^i$. By requiring that the modular transformation of $U^i$ is the same with that of $u^i$, we use the Ansatz that $F$-term of $U^i$, $F^{U_i}$, transform as
\begin{align}
F^{Ui}& ~\to~  \tilde F^{Ui} 
= \frac{\partial \tilde X^i}{\partial X^j} F^{Uj}.
\end{align}

The K\"ahler potential of the moduli $U^i$ is written by
\begin{align}
K_0=-\ln \left[ -i \int_{\cal M} \Omega \wedge \bar \Omega\right]
=-\ln \left[ i |X^0|^2\frac{\kappa_{ijk}}{6} ( U^i -\bar U^i)( U^j  -\bar U^j)( U^k -\bar U^k) \right].
\end{align}
Each of ${\bf \overline{27}}$ matter fields $A^i$ correspond to one of the moduli $U^i$, and 
matter K\"ahler metric is written by\cite{Dixon:1989fj} 
\begin{align}
\hat K_{i \bar j} = e^{-K_0/3}(K_0)_{i\bar j}.
\end{align}
It transforms as 
\begin{align}
K_{i \bar j} ~\to~ \tilde K_{i \bar j} 
=|\tilde X^0|^{2/3} (K)_{\ell \bar m}\frac{\partial  X^\ell}{\partial  \tilde X^i}\frac{\partial {\bar X}^{\bar m}}{\partial \tilde  {\bar X}^{\bar j}}.
\end{align}
Since the K\"ahler potential of matter fields $A^i$ must be invariant, 
the matter fields must transform as 
\begin{align}
A^i ~\to~\tilde A^i =  \frac{\partial \tilde X^i}{\partial X^j}A^j.
\end{align}

The Yukawa coupling is obtained as 
\begin{align}
Y_{ijk}=X_0{\cal F}_{ijk}=\kappa_{ijk}.
\end{align}
It is constant, and transforms as 
\begin{align}
Y_{ijk}~\to~{\tilde Y}_{ijk}=\tilde X^0 \frac{\partial X^\ell}{\partial \tilde X^i}
\frac{\partial X^m}{\partial \tilde X^j}\frac{\partial X^n}{\partial \tilde X^k}{ Y}_{\ell m n}.
\end{align}
That is, the Yukawa coupling behaves as a tensor, which covariantly transforms up to $\tilde X^0$.
Note that the K\"ahler potential $K_0$ transforms as 
\begin{align}
K_0~\to~K_0-\ln |\tilde X^0|^2.
\end{align}
Thus, the superpotential must transform as 
\begin{align}
W~\to~\tilde X^0 W,
\end{align}
for $e^K|W|^2$ to be invariant.

The soft scalar mass terms $m^2_{i \bar j}A^i\bar A^{\bar j}$ can be written in the unnormalized basis in terms of the curvature \cite{Kaplunovsky:1993rd}
\begin{align}
m^2_{i \bar j}=m_{3/2}^2K_{i \bar j} - F^{U\ell } \bar F^{\bar U\bar m}R_{\ell \bar m i \bar j},
\label{eq:m2}
\end{align}
where 
\begin{align}
R_{\ell \bar m i \bar j} &=\partial_\ell \bar \partial_{\bar m}K_{i \bar j} 
-\Gamma^k_{\ell i}K_{k \bar p}\bar \Gamma^{\bar p}_{\bar m \bar j}
\nonumber\\
&= \partial_\ell \bar \partial_{\bar m}K_{i \bar j} 
- K^{n {\bar p}}(\partial_{\ell} K_{i{\bar p}})(\partial_{\bar m} K_{n {\bar j}})
\nonumber\\
&= e^{-K_0/3}\biggl[-\frac{1}{3}(K_0)_{\ell \bar m}(K_0)_{i\bar j} 
+ (\partial_\ell \bar \partial_{\bar m}\partial_i \bar \partial_{\bar j} K_0)
- (K_0)^{n {\bar p}}(\partial_{\ell} \partial_i \partial_{\bar p} K_0)(\partial_{\bar m}\partial_n \partial_{\bar j} K_0)\biggl],
\end{align}
with $\Gamma^k_{\ell n}=K^{k \bar j}\partial_\ell K_{\bar j n}$. 
Since $K_0$ is invariant under the symplectic modular transformation up to the $X^0$ factor, the modular transformation of the Riemann curvature tensor is found to be
\begin{align}
     R_{\ell \bar m i \bar j} \rightarrow  \tilde R_{\ell \bar m i \bar j}
    = |\tilde X^0|^{2/3} 
    \frac{\partial X^p}{\partial \tilde X^\ell}\frac{\partial {\bar X}^{\bar q}}{\partial \tilde {\bar X}^{\bar m}}\frac{\partial X^s}{\partial \tilde X^i}\frac{\partial {\bar X}^{\bar t}}{\partial \tilde {\bar X}^{\bar j}}  R_{p \bar q s \bar t}.
\end{align}
As a result, the transformation of the latter term in Eq. (\ref{eq:m2}) is the same with the K\"ahler metric:
\begin{align}
    F^{U\ell }\bar F^{\bar U\bar m}R_{\ell \bar m i \bar j} \rightarrow \tilde F^{U\ell }\tilde {\bar F}^{\bar U\bar m} \tilde R_{\ell \bar m i \bar j} = |\tilde X^0|^{2/3}  \frac{\partial  X^s}{\partial \tilde X^i}\frac{\partial {\bar X}^{\bar t}}{\partial \tilde{\bar X}^{\bar j}}F^{Up }\bar F^{\bar U\bar q} R_{p \bar q s \bar t}.
\end{align}
In this way, the scalar masses squared $m^2_{i \bar j}$ transforms covariantly under the modular symmetry.

Next, let us discuss the modular transformation of the $A$-term. 
From the general expression of the $A$-term\cite{Kaplunovsky:1993rd},
\begin{align}
h_{ijk} &= F^{U\ell}\left(\partial_{\ell} Y_{ijk}-\Gamma^m_{\ell(i}Y_{jk)m} + \partial_{\ell} K_0 Y_{ijk} \right)
\nonumber\\
&=F^{U\ell}X^0\left(\partial_{\ell}\partial_{i}\partial_{j}\partial_{k} {\cal F}-K^{m\bar n}\partial_{\bar n}\partial_\ell \partial_{(i} K_0\partial_j \partial_{k)}\partial_m {\cal F}+ \partial_{\ell} K_0 \partial_{i}\partial_{j}\partial_{k} {\cal F} \right),
\end{align}
we find that the $A$-term also transforms as
\begin{align}
h_{ijk} &\rightarrow \tilde h_{ijk} =
\tilde{X}^0\frac{\partial  X^{\ell}}{ \partial  \tilde X^i}
\frac{\partial  X^m}{ \partial  \tilde X^j}\frac{\partial  X^n}{ \partial  \tilde X^k}h_{\ell mn},
\end{align}
in the same way as the Yukawa coupling. Thus, 
the $A$-term transforms covariantly under the modular symmetry, and 
the $A$-term coefficient is invariant under the modular symmetry. 

As discussed in the single modulus case, the behavior of $\mu$-term and $B$-term depends on how to generate the $\mu$-term. 
It is interesting to study these behaviors in a concrete model, but we leave it for future work.

\section{Conclusion}
\label{sec:Conclusion}

We have studied the modular symmetry of soft SUSY breaking terms.
Soft scalar masses and $A$-term coefficients are invariant under the modular symmetry, 
when we regard the $F$-term as the spurion with the modular weight $-2$.
The behavior of $\mu$-term and $B$-term depends on how to generate the $\mu$-term.
If the $\mu$-term is generated by the mechanism consistent with the modular symmetry, 
e.g., the Giudice-Masiero mechanism, the $\mu$-term and $B$-terms are invariant.

The flavor structure of sfermion masses including off-diagonal (LR) and (RL) elements is 
determined by $\rho(\gamma)$, which is the same flavor symmetry as Yukawa couplings, i.e., the flavor structure of 
fermion masses.
These results would be important in the effective field theory below the SUSY breaking scale \cite{Kobayashi:2021uam,Kobayashi:2021pav}.
We would study its implications in the standard-model effective field theory.

We also have studied the modular symmetry of soft SUSY breaking terms on CY compactifications in the heterotic string with 
the standard embedding. It can serve as examples with the multi-moduli theory. 
Since our finding expressions are rather general, it would be interesting to study concrete models.

\vspace{1.5 cm}
\noindent
{\large\bf Acknowledgments}\\

This work was supported by JPSP KAKENHI Grant Numbers JP19J00664 (HO), JP20K14477(HO) and 
JP20J20388(HU), and 
JST SPRING Grant Number JPMJSP2119(SK).



\end{document}